\documentclass[aps,preprint,showpacs,nofootinbib]{revtex4}
\usepackage{amsfonts}
\usepackage{amsmath}
\usepackage{amssymb}
\usepackage{relsize}
\usepackage{graphicx}
\usepackage{pdfpages}
\usepackage{subfigure}
\usepackage{float}
\usepackage[caption = false]{subfig}
\usepackage{color}

\usepackage{multirow}
\begin{document}
%\addcolour{Mauve}{0.5}{0.0}{0.5}
%\includegraphics{sears1a.eps}

\title{The exact calculation of the edge components of the angular Fock coefficients}

\author{Evgeny Z. Liverts}
\affiliation{Racah Institute of Physics, The Hebrew University, Jerusalem 91904,
Israel}

%\author{Nir Barnea}
%\affiliation{Racah Institute of Physics, The Hebrew University, Jerusalem 91904,Israel}

%\lyxaddress{Racah Institute of Physics, The Hebrew University, Jerusalem 91904,Israel}
\begin{abstract}
The present paper constitutes the development of our previous work devoted to calculations of the angular Fock coefficients, $\psi_{k,p}(\alpha,\theta)$.
The explicit analytic representations for the edge components  $\psi_{k,0}^{(0)}$ and $\psi_{k,0}^{(k)}$ with $k\leq8$ are derived.
The methods developed enable such a calculation for arbitrary $k$.
The single-series representation for subcomponent $\psi_{3,0}^{(2e)}$ missed in the previous paper is developed.
It is shown how to express some of the complicated subcomponents through the hypergeometric and the elementary functions, as well.
Using the specific operator of the Wolfram \emph{Mathematica}, the simple explicit representations for some complex mathematical objects under consideration are obtained.

\end{abstract}


\maketitle

\section{Introduction}\label{S0}

As far back as the 1935s Bartlett et al \cite{B35} showed that
no ascending power series in the interparticle coordinates $r_1,~r_2$ and $r_{12}$ can be a formal solution of the Schrodinger equation
for the $~^1S$-state of helium. Later Bartlett \cite{B37} argued the existence of the helium ground state expansion included $\ln (r_1^2+r_2^2)$. Finally, Fock \cite{FOCK} proposed the expansion
%The Fock expansion for$~^1S$ states of the two-electron atom/ions is of the form \cite{FOCK}:
\begin{equation}\label{I1}
\Psi(r,\alpha,\theta)=\sum_{k=0}^{\infty}r^k\sum_{p=0}^{[k/2]}\psi_{k,p}(\alpha,\theta)(\ln r)^p,
\end{equation}
where $r=\sqrt{r_1^2+r_2^2}$ denotes the  hyperspherical radius, and the hyperspherical angles $\alpha$ and $\theta$ are defined as
\begin{equation}\label{I2}
\alpha=2\arctan \left(r_2/r_1\right),~~~\theta=\arccos\left[(r_1^2+r_2^2-r_{12}^2)/(2r_1r_2)\right].
\end{equation}
The convergence of expansion (\ref{I1}) for the ground state of helium was rigorously studied in Refs.\cite{MORG,LER}.
The method proposed by Fock \cite{FOCK} for investigating the $~^1S$ helium wave functions was generalized \cite{ES1, DEM} for arbitrary systems of charged particles and for states of any symmetry.
The Fock expansion was somewhat generalized \cite{PLUV} to be applicable to any $S$ state, and its first two terms were determined.
The most comprehensive investigation on the methods of derivation and calculation of the angular Fock coefficients (AFC)
 $\psi_{k,p}(\alpha,\theta)$
was presented in the works of Abbott, Gottschalk and Maslen \cite{AM1,GAM2,GM3}.
In Ref.\cite{LEZ1} the further development of the methods of calculation of the AFC was presented. Separation of the AFC by the components, associated with definite power of the nucleus charge $Z$ was introduced.
Some of the AFC or its components that were not calculated previously, were derived \cite{LEZ1}.

% paper extends and develops the previous paper \cite{LEZ1}.
The present paper improves and develops methods and extends the results obtained in the previous work \cite{LEZ1}.
We derive the exact expressions for the edge components of the most complicated AFC, $\psi_{k,0}(\alpha,\theta)$.
We calculate the subcomponent $\psi_{3,0}^{(2e)}$ missed in \cite{LEZ1}.
We show how to express some of the complicated subcomponents through the elementary functions.
Using the operator \textbf{FindSequenceFunction} of the Wolfram \emph{Mathematica}, we obtain simple explicit representations for some complex mathematical objects under consideration.

%To solve the problems mentioned above, we introduce the main mathematical objects that can serve as the basis for our consideration.
To solve the problems mentioned above, we introduce some mathematical concepts that can serve as a basis for further consideration.
It has been proven that the AFCs  satisfy (see, e.g., \cite{AM1} or \cite{LEZ1}) the Fock recurrence relation (FRR)
\begin{subequations}\label{I4}
\begin{align}
\left[ \Lambda^2-k(k+4)\right]\psi_{k,p}=h_{k,p},~~~~~~~~~~~~~~~~~~~~~~~~~~~~~~~~~~~~~~~~~~~~~~~~~~~~~~~~~~~~~\label{I4a}\\
h_{k,p}=2(k+2)(p+1)\psi_{k,p+1}+(p+1)(p+2)\psi_{k,p+2}-2 V \psi_{k-1,p}+2 E \psi_{k-2,p},\label{I4b}
\end{align}
\end{subequations}
where $E$ is the energy and $V=V_0+Z V_1$ is the dimensionless Coulomb interaction for the two-electron atom/ions.
The electron-electron $V_0$ and the electron-proton $V_1$ interactions are defined as follows
\begin{equation}\label{I5}
V_0=1/\xi,~~~~~~
V_1=-\left[\csc(\alpha/2)+\sec(\alpha/2)\right],
\end{equation}
where the variable
\begin{equation}\label{I3}
\xi=\sqrt{1-\sin \alpha \cos \theta}.
\end{equation}
The hyperspherical angular momentum operator, projected on $S$ states, is:
\begin{equation}\label{I6}
\Lambda^2=-\frac{4}{\sin^2 \alpha}\left(\frac{\partial}{\partial\alpha}\sin^2\alpha\frac{\partial}{\partial\alpha}+\frac{1}{\sin\theta} \frac{\partial}{\partial\theta}\sin\theta \frac{\partial}{\partial\theta}\right),
\end{equation}
and its eigenfunctions are the hyperspherical harmonics (HH)
\begin{equation}\label{I7}
Y_{kl}(\alpha,\theta)=N_{kl}\sin^l\alpha~C_{k/2-l}^{(l+1)}(\cos\alpha)P_l(\cos\theta),~~~~~~~~~~k=0,2,4,...; l=0,1,2,...,k/2
\end{equation}
where the $C_n^\nu(x)$ and $P_l(z)$ are Gegenbauer and Legendre polynomials, respectively. The normalization constant is
\begin{equation}\label{I8}
N_{kl}=2^ll!\sqrt{\frac{(2l+1)(k+2)(k/2-l)!}{2\pi^3(k/2+l+1)!}},
\end{equation}
so that
\begin{equation}\label{I9}
\int Y_{kl}(\alpha,\theta)Y_{k'l'}(\alpha,\theta)d\Omega=\delta_{kk'}\delta_{ll'},
\end{equation}
where $\delta_{mn}$ is the Kronecker delta, and the appropriate volume element is
\begin{equation}\label{I10}
d\Omega=\pi^2 \sin^2\alpha~d\alpha \sin\theta d\theta.~~~~~~~~~~~~~~~~~\alpha\in [0,\pi],~\theta\in [0,\pi]
\end{equation}
It was shown \cite{LEZ1} that any AFC, $\psi_{k,p}$ can be separated into the independent parts (components)
\begin{equation}\label{I11}
\psi_{k,p}(\alpha,\theta)=\sum_{j=p}^{k-p} \psi_{k,p}^{(j)}(\alpha,\theta) Z^j
\end{equation}
 associated with a definite power of $Z$, according to separation of the rhs (\ref{I4b})
\begin{equation}\label{I12}
h_{k,p}(\alpha,\theta)=\sum_{j=p}^{k-p} h_{k,p}^{(j)}(\alpha,\theta) Z^j
\end{equation}
of the FRR.
Accordingly, each of the FRR (\ref{I4}) can be separated into the individual equations (IFRR) for each component:
\begin{equation}\label{I13}
\left[ \Lambda^2-k(k+4)\right]\psi_{k,p}^{(j)}(\alpha,\theta)=h_{k,p}^{(j)}(\alpha,\theta).
\end{equation}

\section{The edge components of the most complicated AFC}\label{S1}

It is well-known that calculations of the logarithmless AFC $\psi_{k,0}(\alpha,\theta)$ with $k=1,2,3,...$ are the most complicated ones.
However, it is easy to show that the edge components $\psi_{k,0}^{(0)}$ and $\psi_{k,0}^{(k)}$ of those AFC can be calculated without any problems.
%It follows from Eqs.(\ref{I11})-(\ref{I12}) for $Z$-separation that the edge components $\psi_{k,p}^{(j)}$ are defined by $Z$-powers $j=p$ and $j=k-p$.
Indeed, for $p=0$ Eq.(\ref{I4b}) reduces to the form
\begin{equation}\label{I14}
h_{k,0}=2(k+2)\psi_{k,1}+2\psi_{k,2}-2V\psi_{k-1,0}+2E\psi_{k-2,0}.
\end{equation}
Substitution of Eqs.(\ref{I11})-(\ref{I12}) for the $Z$-power separation into Eq.(\ref{I14}) yields for the right-hand sides of the IFFR (\ref{I13}) with the mentioned edge components
\begin{equation}\label{I15}
h_{k,0}^{(0)}=-2 V_{0} \psi_{k-1,0}^{(0)} + 2E\psi_{k-2,0}^{(0)},
\end{equation}
\begin{equation}\label{I16}
h_{k,0}^{(k)}=-2V_1 \psi_{k-1,0}^{(k-1)},~~~~~~~~~~~~~~~
\end{equation}
where the angular dependent potentials $V_0$ and $V_1$ are defined by Eq.(\ref{I5}).
It is seen that the rhs (\ref{I15}) and (\ref{I16}) of the order $k$ are represented by the corresponding edge components of the order $k-1$ and $k-2$ (if exists).
Moreover, taking into account that $\psi_{1,0}^{(1)}$ is a function of $\alpha$ (see Eq.(49)\cite{LEZ1}), and $\psi_{1,0}^{(0)}$ is a function of $\xi$ (see Eq.(36)\cite{LEZ1}), one can conclude that any component $\psi_{k,0}^{(k)}$ must be a function of a single angle $\alpha$, whereas  $\psi_{k,0}^{(0)}$ must be a function of a single variable $\xi$ defined by Eq.(\ref{I3}). Representation (\ref{I5}) for the potentials was certainly used to make the above conclusions.

There is a specific difference between derivations of the edge components $\psi_{k,0}^{(0)}$, $\psi_{k,0}^{(k)}$ for even and odd $k$. Therefore, we shall present such calculations in details for $k=4$ and $k=5$, whereas for $k=6,7,8$ the corresponding results will be presented without derivations.

%\subsection{Derivation of the AFC components $\psi_{4,0}^{(0)}$ and $\psi_{4,0}^{(4)}$ }\label{S1a}

The general IFRR (\ref{I13}) for $k=4,~p=0,~j=0$ reduces to the form
\begin{equation}\label{1}
\left( \Lambda^2-32\right)\psi_{4,0}^{(0)}=h_{4,0}^{(0)},
\end{equation}
where
\begin{equation}\label{2}
h_{4,0}^{(0)}\equiv -2V_0 \psi_{3,0}^{(0)}+2E \psi_{2,0}^{(0)}=
-\frac{1}{36}\left[\xi^2(E-2)+3(1-2E)^2\right]
\end{equation}
according to relation (\ref{I15}).
The components $\psi_{3,0}^{(0)}$ and $ \psi_{2,0}^{(0)}$ are presented in Table I and Eq.(55) of Ref.\cite{LEZ1}, respectively.
%$E$ is the energy of the two-electron ion/atom in $S$ state.
It is seen that the rhs (\ref{2}) of the IFRR (\ref{1}) is a function of a single variable $\xi$.
It was shown in \cite{LEZ1} that in this case the solution of the IFRR (\ref{I13}) with the rhs $h_{k,p}^{(j)}(\alpha,\theta)\equiv \textsl{h}(\xi)$ reduces to solution of the differential equation
\begin{equation}\label{3}
\left(\xi^2-2\right)\Phi_k''(\xi)+\frac{5\xi^2-4}{\xi}\Phi_k'(\xi)-k(k+4)\Phi_k(\xi)=\textsl{h}(\xi),
\end{equation}
where $\Phi_k(\xi)\equiv \psi_{k,p}^{(j)}(\alpha,\theta)$. The particular solution $\Phi_k^{(p)}$ of Eq.(\ref{3}) can be found by the method of variation of parameters in the form
\begin{equation}\label{4}
\Phi_k^{(p)}(\xi)=\frac{1}{(k+2)\sqrt{2}}\left[
\textsl{u}_k(\xi)\int\textsl{v}_k(\xi)f(\xi)d\xi-
\textsl{v}_k(\xi)\int\textsl{u}_k(\xi)f(\xi)d\xi\right],
\end{equation}
where $f(\xi)=\textsl{h}(\xi)\xi^2\sqrt{2-\xi^2}$. The linearly independent solutions of the homogeneous equation associated with Eq.(\ref{3}) are defined by
\begin{equation}\label{5}
 \textsl{u}_k(\xi)=\frac{P_{k+3/2}^{1/2}\left(\xi/\sqrt{2}\right)}{\xi\sqrt[4]{2-\xi^2}},~~~~~~~~~~~~
\textsl{v}_k(\xi) =\frac{Q_{k+3/2}^{1/2}\left(\xi/\sqrt{2}\right)}{\xi\sqrt[4]{2-\xi^2}},
\end{equation}
where  $P_\nu^\mu(x)$ and $Q_\nu^\mu(x)$ are the associated Legendre functions of the first and second kind, respectively.
The general solution of the inhomogeneous equation (\ref{3}) is certainly of the form
\begin{equation}\label{6}
\Phi_k^{(p)}(\xi)+c_{1,k} \textsl{u}_k(\xi)+c_{2,k} \textsl{v}_k(\xi),
\end{equation}
where the values of coefficients $c_{1,k}$ and $c_{2,k}$ are defined by requirements of the finiteness and "purity" of the final physical solution.
The first requirement means that any component $\psi_{k,p}^{(j)}(\alpha,\theta)$ of the AFC must be finite at each point of the two-dimensional angular space described by the hyperspherical angles $\alpha\in[0,\pi]$ and $\theta\in[0,\pi]$. The second requirement associates with even values of $k$ (only) and concerns the obtaining of the single-valued solution containing no admixture of the HH $Y_{kl}(\alpha,\theta)$.

Turning to the component $\psi_{4,0}^{(0)}$, and simplifying the solutions (\ref{5}) for $k=4$, one obtains
\begin{equation}\label{7}
 \textsl{u}_4(\xi)=\frac{2^{3/4}\left[\xi^2(3-2\xi^2)^2-1\right]}{\xi\sqrt{\pi(2-\xi^2)}},~~~~~
 \textsl{v}_4(\xi)=-\frac{\sqrt{\pi}}{2^{1/4}}\left(4\xi^4-8\xi^2+3\right).
\end{equation}
Substitution of the representations (\ref{2}) and (\ref{7}) into (\ref{4}) yields
\begin{equation}\label{8}
\Phi_4^{(p)}(\xi)=\frac{\xi^2}{1440}\left[10(2E-1)^2-\xi^2(20 E^2-21 E+7)\right].
\end{equation}
It is seen that the particular solution $\Phi_4^{(p)}$, as well as $\textsl{v}_4(\xi)$, are regular over the relevant angular space, whereas
$\textsl{u}_4(\xi)$ is singular at the points $\xi=0~(\alpha=\pi/2,\theta=0)$ and $\xi=\sqrt{2}~(\alpha=\pi/2,\theta=\pi)$.
Hence, first of all, one should set $c_{1,4}=0$ in Eq.(\ref{6}) in order to comply with the finiteness condition.
It is clear that the requirement of "purity" reduces to the orthogonality condition
\begin{equation}\label{9}
\int\psi_{4,0}^{(0)}(\alpha,\theta)Y_{4l}(\alpha,\theta)d\Omega=0.
\end{equation}
Given that $\psi_{4,0}^{(0)}=\Phi_4^{(p)}(\xi)+c_{2,4}\textsl{v}_4(\xi)$, one obtains for the coefficient
\begin{equation}\label{10}
c_{2,4}=-\int\Phi_4^{(p)}(\xi)Y_{40}(\alpha,\theta)d\Omega\left(\int\textsl{v}_4(\xi)Y_{40}(\alpha,\theta)d\Omega\right)^{-1}=
\frac{E(21-20E)-7}{2880\sqrt{\pi}~2^{3/4}}.
\end{equation}
Whence, the final result for the "pure" component is
%one finally obtains the "pure" component in the form
\begin{equation}\label{10a}
\psi_{4,0}^{(0)}=\frac{60E^2-63E+21+8\xi^2(E-2)}{5760}.
\end{equation}
Note that in order to derive Eq.(\ref{10}) we put $l=0$ in Eq.(\ref{9}). However, putting $l=2$ one obtains the same result, whereas for $l=1$ one obtains identity.

Next step is deriving the solution of the IFRR (\ref{I13}) for $k=4,~p=0,~j=4$ which becomes
\begin{equation}\label{11}
\left( \Lambda^2-32\right)\psi_{4,0}^{(4)}=h_{4,0}^{(4)}.
\end{equation}
Expression (\ref{I16}) yields for the rhs of Eq.(\ref{11})
\begin{equation}\label{12}
h_{4,0}^{(4)}\equiv -2V_1 \psi_{3,0}^{(3)}=-
\frac{1}{18}(2+5 \sin \alpha)\left[\tan\left(\frac{\alpha}{2}\right)+\cot\left(\frac{\alpha}{2}\right)+2\right],
\end{equation}
where the component $\psi_{3,0}^{(3)}$ is presented in Table I of Ref.\cite{LEZ1}.
Turning to the variable
\begin{equation}\label{13}
\rho=\tan(\alpha/2),
\end{equation}
one obtains
\begin{equation}\label{14}
h(\rho)\equiv h_{4,0}^{(4)}(\alpha,\theta)=-\frac{(1+\rho)^2(1+5\rho+\rho^2)}{9\rho(1+\rho^2)}.
\end{equation}
It was shown (see Sec.V in Ref.\cite{LEZ1}) that in case of the rhs $h_{k,p}^{(j)}$ of the IFRR (\ref{I13}) reduces to the function $\textmd{h}(\alpha)\equiv h(\rho)$ of a single variable $\alpha$ (or $\rho$), the solution of Eq.(\ref{I13}) represents a function $g(\rho)\equiv \psi_{k,p}^{(j)}(\alpha)$ satisfying the differential equation
\begin{equation}\label{15}
\left(1+\rho^2\right)^2g''(\rho)+2\rho^{-1}\left(1+\rho^2\right)g'(\rho)+k(k+4)g(\rho)=-h(\rho).
\end{equation}
Method of variation of parameters enables us to obtain the particular solution of Eq.(\ref{15}) in the form
%Eq.(45)\cite{LEZ1} yields for the particular solution of Eq.(\ref{11}) with the rhs defined by Eq.(\ref{14}),
\begin{equation}\label{16}
g(\rho)=u_{k}(\rho)\int \frac{v_{k}(\rho)h(\rho)\rho^2}{(\rho^2+1)^3}d\rho-
v_{k}(\rho)\int \frac{u_{k}(\rho)h(\rho)\rho^2}{(\rho^2+1)^3}d\rho,
\end{equation}
where the linearly independent solutions of the homogeneous equation associated with Eq.(\ref{15}) are
\begin{equation}\label{17}
u_{k}(\rho)=\frac{(1+\rho^2)^{k/2+2}}{\rho}~_2F_1\left(\frac{k+3}{2},\frac{k}{2}+1,\frac{1}{2};-\rho^2\right),
\end{equation}
\begin{equation}\label{18}
v_{k}(\rho)=(1+\rho^2)^{k/2+2}~_2F_1\left(\frac{k+3}{2},\frac{k}{2}+2,\frac{3}{2};-\rho^2\right).
\end{equation}
The Gauss hypergeometric function $~_2F_1$ was introduced in Eqs.(\ref{17}),(\ref{18}).
The general solution of the inhomogeneous equation (\ref{15}) is defined as
\begin{equation}\label{18a}
g(\rho)+b_{1,k}u_k(\rho)+b_{2,k}v_k(\rho),
\end{equation}
where the coefficients $b_{1,k}$ and $b_{2,k}$ can be determined by the requirements of finiteness and "purity"
as it was explained earlier.
Turning to the considered case of $k=4$ , one obtains for the independent solutions of the homogeneous equation:
\begin{equation}\label{19}
u_{4}(\rho)=\frac{(1+\rho^2)^4}{\rho}~_2F_1\left(\frac{7}{2},3,\frac{1}{2};-\rho^2\right)=
\frac{(1-\rho^2)(1-4\rho+\rho^2)(1+4\rho+\rho^2)}{\rho(1+\rho^2)^2},
\end{equation}
\begin{equation}\label{20}
v_{4}(\rho)=(1+\rho^2)^4~_2F_1\left(\frac{7}{2},4,\frac{3}{2};-\rho^2\right)=
\frac{(\rho^2-3)(3\rho^2-1)}{3(1+\rho^2)^2}.~~~~~~~~~~~~~~~~~~~~~~~
\end{equation}
Substitution of the representations (\ref{19}), (\ref{20}) and (\ref{14}) into the rhs of Eq.(\ref{16}) yields for the particular solution
\begin{equation}\label{21}
\psi_{4,0}^{(4p)}=\frac{\rho(3+7\rho+3\rho^2)}{54(1+\rho^2)^2}=
\frac{1}{216}(6+7\sin \alpha) \sin \alpha.
\end{equation}
It is seen that the particular solution $g(\rho)$ and the solution $v_4(\rho)$ of the homogeneous equation are regular over the relevant angular space, whereas
$u_4(\rho)$ is singular at the points $\rho=0~(\alpha=0)$.
Hence, one should set $b_{1,4}=0$ in Eq.(\ref{18a}) in order to comply with the finiteness condition.
The requirement of "purity" can be expressed through the relation
\begin{equation}\label{23}
\int\psi_{4,0}^{(4)}(\alpha,\theta)Y_{4l}(\alpha,\theta)d\Omega=0.
\end{equation}
Given that $\psi_{4,0}^{(4)}=g(\rho)+b_{2,4}v_4(\rho)$, one obtains for the coefficient
\begin{equation}\label{24}
b_{2,4}=-\int g(\rho)Y_{40}(\alpha,\theta)d\Omega\left(\int v_4(\rho)Y_{40}(\alpha,\theta)d\Omega\right)^{-1}=
\frac{7}{288}+\frac{2}{45\pi}.
\end{equation}
Whence, the final result for the "pure" component reads
\begin{equation}\label{25}
\psi_{4,0}^{(4)}=\frac{120 \pi \sin \alpha+128 \cos(2\alpha)+105\pi+64}{4320\pi}.
\end{equation}
Note that for $l=1,2$  the orthogonality condition (\ref{23}) represents identity.

Putting $k=5,~p=0,~j=0$ in Eq.(\ref{I13}), one obtains
\begin{equation}\label{43}
\left( \Lambda^2-45\right)\psi_{5,0}^{(0)}=h_{5,0}^{(0)},
\end{equation}
where according to relation (\ref{I15})
\begin{equation}\label{44}
h_{5,0}^{(0)}\equiv -2V_0 \psi_{4,0}^{(0)}+2E \psi_{3,0}^{(0)}=
\frac{63 E-21-60E^2+8(2+29E-60E^2)\xi^2+80E(E-2)\xi^4}{2880\xi}.
\end{equation}
We used Eq.(\ref{10a}) for representation of the component $\psi_{4,0}^{(0)}$.

The independent solutions (\ref{5}) of the homogeneous equation associated with Eq.(\ref{3}), for $k=5$ become
\begin{equation}\label{45}
 \textsl{u}_5(\xi)=\frac{2^{1/4}(8\xi^6-28\xi^4+28\xi^2-7)}{\sqrt{\pi(2-\xi^2)}},~~~~~~~
\textsl{v}_5(\xi) =\frac{\sqrt{\pi}(1-12\xi^2+20\xi^4-8\xi^6)}{2^{3/4}\xi}.
\end{equation}
Substitution of the representations (\ref{44}) and (\ref{45}) into the particular solution (\ref{4}) of the inhomogeneous equation (\ref{3}) for $k=5$ yields
\begin{equation}\label{46}
\Phi_5^{(p)}=\frac{\xi}{172800}
\left[45(7-21E+20E^2)-5(113-199E+60E^2)\xi^2+2(113-119E+20E^2)\xi^4\right].
\end{equation}
It is seen that the particular solution (\ref{46}) is regular over the relevant angular space, whereas
$\textsl{u}_5(\xi)$ is singular at the point $\xi=\sqrt{2}$, and $\textsl{v}_5(\xi)$ is singular at the point $\xi=0$. Hence, the physical solution of Eq.(\ref{43}) is coincident with the partial solution (\ref{46}), that is
\begin{equation}\label{47}
\psi_{5,0}^{(0)}=\Phi_5^{(p)}.
\end{equation}

The general IFRR (\ref{I13}) for $k=5,~p=0,~j=5$ reduces to the form
\begin{equation}\label{48}
\left( \Lambda^2-45\right)\psi_{5,0}^{(5)}=h_{5,0}^{(5)},
\end{equation}
where according to Eq.(\ref{I16})
\begin{equation}\label{49}
h_{5,0}^{(5)}\equiv -2V_1 \psi_{4,0}^{(4)}=
\frac{(1+\rho)\left[64(3-10\rho^2+3\rho^4)+15\pi(1+\rho^2)(7+16\rho+7\rho^2)\right]}
{2160\pi \rho(1+\rho^2)^{3/2}},
\end{equation}
expressed through the variable $\rho$ defined by Eq.(\ref{13}).
According to Eqs.(\ref{17}), (\ref{18}) the linearly independent solutions of the homogeneous equation associated with Eq.(\ref{15}) for $k=5$ can be simplified to the form
\begin{equation}\label{50}
u_{5}(\rho)=\frac{(1+\rho^2)^{9/2}}{\rho}~_2F_1\left(4,\frac{7}{2},\frac{1}{2};-\rho^2\right)=
\frac{1-7\rho^2(3-5 \rho^2+\rho^4)}{\rho(1+\rho^2)^{5/2}},
\end{equation}
\begin{equation}\label{51}
v_{5}(\rho)=(1+\rho^2)^{9/2}~_2F_1\left(4,\frac{9}{2},\frac{3}{2};-\rho^2\right)=
\frac{7-35\rho^2+21\rho^4-\rho^6}{7(1+\rho^2)^{5/2}}.
\end{equation}
Substituting representations (\ref{49})-(\ref{51}) into the rhs of Eq.(\ref{16}), one obtains
for the particular solution of Eq.(\ref{15}) with $k=5$
\begin{eqnarray}\label{52}
\psi_{5,0}^{(5p)}=-\frac{1}{453600\pi \rho(1+\rho^2)^{5/2}}
\left[64(23+161\rho-168\rho^2-700\rho^3+105\rho^4+315\rho^5)+
\right.
\nonumber~~~\\
\left.
15\pi(43+301\rho-168\rho^2-700\rho^3+805\rho^4+735\rho^5)\right]
~~~.
\end{eqnarray}
The physical solution $\psi_{5,0}^{(5)}$ of the IFRR (\ref{48}) must be finite for all values of $0\leq\alpha\leq\pi$ and hence for $\rho\geq0$.
Therefore, let us consider the power series expansions of the particular solution $\psi_{5,0}^{(5p)}(\rho)$ and the individual solutions $u_5(\rho)$ and $v_5(\rho)$ of the corresponding homogeneous equation about $\rho=0$ and $\rho=\infty$. One obtains:
\begin{subequations}\label{53}
\begin{align}
\psi_{5,0}^{(5p)}(\rho)\underset{\rho\rightarrow 0}{=}
-\frac{1}{\rho}\left(\frac{43}{30240}+\frac{46}{14175\pi}\right)-\left(\frac{43}{4320}+\frac{46}{2025\pi}\right)+
O(\rho),~~~~~~~~~~~~~~~~~\label{53a}\\
\psi_{5,0}^{(5p)}(\rho)\underset{\rho\rightarrow \infty}{=}
-\frac{1}{\rho}\left(\frac{7}{288}+\frac{2}{45\pi}\right)-\frac{1}{\rho^2}\left(\frac{23}{864}+
\frac{2}{135\pi}\right)+O\left(\frac{1}{\rho^3}\right)
.~~~~~~~~~~~~~~~~~~\label{53b}
\end{align}
\end{subequations}
\begin{subequations}\label{54}
\begin{align}
u_5(\rho)\underset{\rho\rightarrow 0}{=}
\frac{1}{\rho}-\frac{47\rho}{2}+O(\rho^3),~~~~~~~~~~~~~~~~~~~~~~~~~~~~~\label{54a}\\
u_5(\rho)\underset{\rho\rightarrow \infty}{=}-7+\frac{105}{2\rho^2}+O\left(\frac{1}{\rho^3}\right)
.~~~~~~~~~~~~~~~~~~~~~~~\label{54b}
\end{align}
\end{subequations}
\begin{subequations}\label{55}
\begin{align}
v_5(\rho)\underset{\rho\rightarrow 0}{=}
1-\frac{15\rho^2}{2}+O(\rho^3),~~~~~~~~~~~~~~~~~~~~~~~~~~~~~\label{55a}\\
v_5(\rho)\underset{\rho\rightarrow \infty}{=}-\frac{\rho}{7}+\frac{47}{14\rho}+O\left(\frac{1}{\rho^3}\right)
.~~~~~~~~~~~~~~~~~~~~~~~\label{55b}
\end{align}
\end{subequations}
It is seen that $v_5(\rho)$ is divergent as $\rho\rightarrow \infty$, whereas $u_5(\rho)$ and $\psi_{5,0}^{(5p)}(\rho)$ are singular at the point $\rho=0$.
Thus, in order to comply with the finiteness condition, one should set $b_{2,5}=0$ and
\begin{equation*}
b_{1,5}=\frac{43}{30240}+\frac{46}{14175\pi}
\end{equation*}
 in the general solution
\begin{equation*}
\psi_{5,0}^{(5p)}(\rho)+b_{1,5}u_5(\rho)+b_{2,5}v_5(\rho).
\end{equation*}
The final result expressed in terms of the hyperspherical angle $\alpha$ is
\begin{equation}\label{56}
\psi_{5,0}^{(5)}=-\frac{\left[\cos\left(\frac{\alpha}{2}\right)+\sin\left(\frac{\alpha}{2}\right)\right]
\left[4(32+45\pi)+3(448+155\pi)\cos(2\alpha)+(704+465\pi)\sin \alpha
\right]}{64800\pi}.
\end{equation}
It is clear that using the technique described above, one can subsequently calculate the edge components of any given order $k$.
Here we present such components up to $k=8$. They are
\begin{eqnarray}\label{57}
\psi_{6,0}^{(0)}=\frac{1}{29030400}
\left[3007-11361E+16460E^2-10080E^3-
\right.
\nonumber~~~~~~~~~~~~~~~~~~~~~~~~~~~~~~~~~~~~~\\
\left.
(4180E^2-12705E+6215)\xi^2+24(113-119E+20E^2)\xi^4\right],~~~~~~~~~~~~
\end{eqnarray}
\begin{eqnarray}\label{58}
\psi_{7,0}^{(0)}=\frac{\xi}{36578304000}\left[630(3007-11361E+16460E^2-10080E^3)+
\right.
~\nonumber~~~~~~~~~~~~~~~~~~~~~~~~~\\
\left.
105(30240E^3-141700E^2+164283E-60341)\xi^2-
\right.
\nonumber~~~~~~~~~~~~~~~~~~~~~~~~~~~\\
\left.
14(60480E^3-491860E^2+873789E-430523)\xi^4+
\right.
\nonumber~~~~~~~~~~~~~~~~~~~~~~~~~~~\\
\left.
4(22680E^3-266950E^2+660219E-430523)\xi^6\right],~~~~~~~~~~~~~~~~~~~~~~~~~~~
\end{eqnarray}
\begin{eqnarray}\label{59}
\psi_{8,0}^{(0)}=\frac{1}{21069103104000}\left[5(30481920E^4-68266800E^3+72613544E^2-39544113E+8871475)+
\right.
~\nonumber~\\
\left.
40(3250800E^3-13370360E^2+13273113E-4324891)\xi^2-
\right.
\nonumber~~~~~~~~~~~~~~~~~~~\\
\left.
420(71280E^3-556120E^2+934809E-430523)\xi^4+
\right.
\nonumber~~~~~~~~~~~~~~~~~~~\\
\left.
128(22680E^3-266950E^2+660219E-430523)\xi^6\right],~~~~~~~~~~~~~~~~~~~
\end{eqnarray}
\begin{eqnarray}\label{60}
\psi_{6,0}^{(6)}=\frac{1}{21772800\pi}
\left[80(448+255\pi)+144(448+155\pi)\cos(2\alpha)+
\right.
\nonumber~~~~~~~~~~~~~~~~~~~~~~~~~~~~~~~~~~~~~\\
\left.
315(64+55\pi)\sin \alpha+7(10816+4335\pi)\sin(3\alpha)\right],~~~~~~~~~~~~
\end{eqnarray}
\begin{eqnarray}\label{61}
\psi_{7,0}^{(7)}=-\frac{\left[\cos\left(\frac{\alpha}{2}\right)+\sin\left(\frac{\alpha}{2}\right)\right]}{609638400\pi}
\left[3(585\pi-10304)\sin \alpha+3(58845\pi+142016)\sin(3\alpha)+
\right.
\nonumber~~~~~~~~~~~\\
\left.
8(4485\pi+12992)\cos(2\alpha)+97560\pi+211456\right],~~~~~~~~~~~~~
\end{eqnarray}
\begin{eqnarray}\label{62}
\psi_{8,0}^{(8)}=\frac{1}{1843546521600\pi^2}
\left\{94502912+75\pi(1946944+626787\pi)+
\right.
\nonumber~~~~~~~~~~~~~~~~~~~~~~~~~~~~~~~~~~~~~~~~~~~\\
\left.
2\left[94502912+3\pi(43273408+12251925\pi)\right]\cos(2\alpha)+4096(46144+19275\pi)\cos(4\alpha)+
\right.
\nonumber~~~~~~~~\\
1008\pi\left[105(448+225\pi)\sin \alpha+(142016+58845\pi)\sin(3\alpha)\right]
\left.
\right\}.~~~~~~~~~~~~~~~~~~~~
\end{eqnarray}

\section{Single-series representation for subcomponent $\psi_{3,0}^{(2e)}$}\label{S2}

It follows from Eqs.(79),(85) of Ref.\cite{LEZ1} that subcomponent $\psi_{3,0}^{(2e)}$ corresponding to the rhs
\begin{equation}\label{63}
h_{3,0}^{(2e)}=-\frac{\sin \alpha}{\xi},
\end{equation}
of the IFFR
\begin{equation}\label{64}
(\Lambda^2-21)\psi_{3,0}^{(2e)}=h_{3,0}^{(2e)}
\end{equation}
was missed in Ref.\cite{LEZ1}.
Using the technique described in Sec.V of Ref.\cite{LEZ1}, we have found
the mentioned subcomponent (details can be found in Appendix \ref{SA}) in the form of the single-series representation
\begin{equation}\label{65}
\psi_{3,0}^{(2e)}=\sum_{l=0}^\infty P_l(\cos \theta)(\sin \alpha)^l \lambda_l(\rho).
\end{equation}
For $0\leq\rho\leq1$ functions $\lambda_l(\rho)$ can be written in the form
\begin{equation}\label{66}
\lambda_l(\rho)=\frac{1}{2l+1}\left\{u_{3l}(\rho)\mathcal{V}_{3l}(\rho)-v_{3l}(\rho)
\left[\mathcal{U}_{3l}(\rho)-(2l+1)s_l\right]\right\},
\end{equation}
where
\begin{equation}\label{67}
u_{3l}(\rho)=\frac{\left(\rho^2+1\right)^{l-\frac{3}{2}}}{\rho^{2l+1}}
\left[\frac{(2l+3)(2l+5)}{(2l-3)(2l-1)}\rho^4+\frac{2(2l+5)}{2l-1}\rho^2+1\right],
\end{equation}
\begin{equation}\label{68}
v_{3l}(\rho)=\left(\rho^2+1\right)^{l-\frac{3}{2}}
\left[\frac{(2l-3)(2l-1)}{(2l+3)(2l+5)}\rho^4+\frac{2(2l-3)}{2l+3}\rho^2+1\right],
\end{equation}
\begin{equation}\label{69}
\mathcal{U}_{3l}(\rho)=
-\frac{l(l+1)\left[(\rho^2+1)^4\arctan (\rho)+\rho^7-\rho\right]+(l^2-7l-10)\rho^5-(l^2+9l-2)\rho^3}
{2^l(2l-3)(2l-1)(\rho^2+1)^4},
\end{equation}
\begin{eqnarray}\label{70}
\mathcal{V}_{3l}(\rho)=
~\nonumber~~~~~~~~~~~~~~~~~~~~~~~~~~~~~~~~~~~~~~~~~~~~~~~~~~~~~~~~~~~~~~~~~~~~~~~~~~~~~~~~~~~~~~~~~~~~~~~~~~~~~~~\\
-\left[(-2)^l(l-2)(l-1)(2l+3)(2l+5)\right]^{-1}
%\left\{12-(l-3)(2l-3)\rho^2\left[2l^2+l-7+(l-2)(2l-1)\rho^2\right]B_{-\rho^2}(l+1,-3)-
\left\{12\left[B_{-\rho^2}(l+1,-3)-B_{-\rho^2}(l+1,-4)\right]+
\right.
~~~\nonumber~\\
\left.
(2l-3)\rho^2\left[2l^2+l-7+(l-2)(2l-1)\rho^2\right]\left[(3-l)B_{-\rho^2}(l+1,-3)-4B_{-\rho^2}(l+1,-4)\right]
\right\}
.~~~~
\end{eqnarray}
Here $B_z(a,b)$ is the Euler beta function.
It is seen that expression (\ref{70}) cannot be applied directly for $l=1,2$. For this values of $l$, one obtains
\begin{equation}\label{71}
\mathcal{V}_{31}(\rho)=
\frac{1}{140}\left[\frac{3+10\rho^2+11\rho^4-20\rho^6}{(1+\rho^2)^4}+2\ln(1+\rho^2)\right],~~~~~~~~~~
\end{equation}
\begin{equation}\label{72}
\mathcal{V}_{32}(\rho)=
-\frac{1}{84}\left[\frac{5+14\rho^2+9\rho^4-6\rho^6+24\rho^8+6\rho^{10}}{6(1+\rho^2)^4}+\ln(1+\rho^2)\right].
\end{equation}
Using the \emph{Mathematica} operator \textbf{FindSequenceFunction} we have found a simple representation (details can be found in Appendix \ref{SA}) for the coefficient
\begin{equation}\label{73}
s_l=\frac{2^{-l-3}}{(2l-3)(2l-1)(2l+1)}\left[2l(l+1)\left(H_{\frac{l+1}{2}}-H_{\frac{l}{2}}-\pi\right)+2l+3\right\},
\end{equation}
being a part of expression (\ref{66}) for $\lambda_l(\rho)$. Functions $H_z$ give the harmonic numbers.
Remind that for $\rho>1$ one should replace $\rho$ by $1/\rho$ in Eqs.(\ref{66})-(\ref{72}).

\section{Elaboration of some results obtained previously}\label{S3}

In paper \cite{LEZ1} the various components of the AFC were derived in the form of the one-dimensional series with fast convergence.
In particular, the solution of the IFRR
\begin{equation}\label{90}
\left(\Lambda^2-32\right)\psi_{4,1}^{(2d)}=h_{4,1}^{(2d)},~~~~~~~~~~~~~~~~~
\end{equation}
with the rather complicated rhs
\begin{equation}\label{91}
h_{4,1}^{(2d)}=\frac{\pi-2}{3\pi}
\left[\sin\left(\frac{\alpha}{2}\right)+\cos\left(\frac{\alpha}{2}\right)  \right]\left[\frac{5}{3\sin \alpha}\xi^3+\left(1-\frac{2}{\sin\alpha}\right)\xi-\frac{1}{\xi}\right],~~
\end{equation}
was represented by single series of the form
\begin{equation}\label{92}
\psi_{4,1}^{(2d)}=\sum_{l=0}^\infty P_l(\cos\theta)(\sin \alpha)^l \tau_l(\rho),
\end{equation}
where the variables $\xi$ and $\rho$ are defined by Eqs.(\ref{I3}) and (\ref{13}), respectively.
It was shown that for $l>2$  function $\tau_l(\rho)$ can be expressed by the formula
\begin{equation}\label{93}
\tau_l(\rho)=\tau_l^{(p)}(\rho)+A_2(l) v_{4l}(\rho),
\end{equation}
where
\begin{equation}\label{94}
\tau_l^{(p)}(\rho)=\frac{1}{2l+1}\left[u_{4l}(\rho)\mathcal{V}_{4l}(\rho)-v_{4l}(\rho)\mathcal{U}_{4l}(\rho)\right],
\end{equation}
\begin{subequations}\label{95}
\begin{align}
u_{4l}(\rho)=\rho^{-2l-1}(\rho^2+1)^{l+4}~_2F_1\left(\frac{7}{2},3-l;\frac{1}{2}-l;-\rho^2\right),~~~~~~~\label{95a}\\
v_{4l}(\rho)=(\rho^2+1)^{l+4}~_2F_1\left(\frac{7}{2},4+l;l+\frac{3}{2};-\rho^2\right).~~~~~~~~~~~~~~~~\label{95b}
\end{align}
\end{subequations}
However, function $\tau_l^{(p)}(\rho)$ as well as the coefficient $A_2(l)$ were represented in the closed form (see, Eq.(C14)\cite{LEZ1}) only for given  $l\leq10$.

Here, we present the mentioned above functions in a few general closed forms which applicable for any $l\geq3$.
In particular, it is shown in Appendix \ref{SB} that the functions $\mathcal{U}_{4l}$ and $\mathcal{V}_{4l}$ included into the rhs of expression (\ref{94}) can be represented in the form:
\begin{eqnarray}\label{96}
\mathcal{U}_{4l}(\rho)=a_{0l}\frac{8(l-3)!}{15\sqrt{\pi}\Gamma(l+1/2)}
\sum_{m=0}^{l-3}\frac{\Gamma(m+7/2)\Gamma(l-m+1/2)(-1)^m}{m!(l-m-3)!}\times
~~\nonumber~~~~~~~~~~~~~~~~~~~~~~\\
\sum_{n=1}^5 a_{nl}\left(\frac{\rho^{2m+n}-1}{2m+n}+\frac{\rho^{2m+n+1}-1}{2m+n+1}\right).~~~~~~~~~~~~~~~~
\end{eqnarray}
\begin{eqnarray}\label{97}
\mathcal{V}_{4l}(\rho)=\frac{4a_{0l}\Gamma(l+3/2)}{15\sqrt{\pi}}\rho^{2l+3}
\sum_{m=0}^2 \frac{m!}{\left(1+\rho^2\right)^{5-m}}\sum_{k=0}^m \frac{\Gamma(k+7/2)}{k!(m-k)!\Gamma(k+l+3/2)}
\left(-\frac{\rho^2}{1+\rho^2}\right)^k
%\frac{\left(-\rho^2\right)^k} {\left(1+\rho^2\right)^{k}}
\times
~~\nonumber~~~~~~\\
\left[\frac{(k+l+3)!\Gamma(l+m+3/2)}{(l+3)!\Gamma(k+l+m+5/2)}b_{2m+1,l}\rho^{2m}~
_2F_1\left(l+m-1,m-\frac{3}{2};k+l+m+\frac{5}{2};-\rho^2\right)
\right.
~~\nonumber~~~~~~\\
\left.
+\frac{b_{2(2-m),l}}{k+l-m+3}\rho^{3-2m}~_2F_1\left(l-2,m-\frac{3}{2};k+l+\frac{3}{2};-\rho^2\right)\right],~~~~~~~~~~~~~~~
\end{eqnarray}
where
\begin{eqnarray}\label{98}
a_{0l}=-\frac{(\pi-2)2^{-l-1}}{3\pi(2l-1)(2l+3)},~~
a_{1l}=\frac{15-4l(l+1)(4l+11)}{(2l-3)(2l+5)},~~
a_{2l}=4l(2l+3),~~
~~\nonumber~~~~~~\\
a_{3l}=2,~~
a_{4l}=4(l+1)(2l-1),~~
a_{5l}=\frac{(2l-1)(4l+5)}{2l+5},~~~~~~~~~~~~~~~~~~~~~~~~
\end{eqnarray}
\begin{equation}\label{99}
b_{0,l}=a_{1l},~~b_{5,l}=a_{5l},~~b_{s,l}=a_{sl}+a_{s+1l}~~~(s=1,2,3,4).
\end{equation}
Moreover, it is shown in Appendix \ref{SB} that all of the Gauss hypergeometric functions are contained in Eqs.(\ref{95}) and (\ref{97}) can be expressed through the elementary functions. A simple representation of the function $\mathcal{V}_{4l}(\rho)$ through the generalized hypergeometric functions $~_3F_2$ is derived too.

Making use of the \emph{Mathematica} operator \textbf{FindSequenceFunction}, the following representation was derived (see details in Appendix \ref{SC}) for the factor $A_2(l)$
being a part of the rhs of Eq.(\ref{93}):
\begin{eqnarray}\label{100}
A_2(l)=\frac{(2-\pi)\pi^{-3/2}}{360 l (l-2)\Gamma(l+\frac{1}{2})}\times
~\nonumber~~~~~~~~~~~~~~~~~~~~~~~~~~~~~~~~~~~~~~~~~\\
\left\{
\frac{\left[l(l+1)(688 l^4+1376 l^3-2480 l^2-3168 l+465)+450\right]\Gamma\left(\frac{l-1}{2}\right)\Gamma\left(\frac{l+1}{2}\right)}
{(2l-3)(2l-1)(2l+1)(2l+3)(2l+5)}-\frac{56}{l-1}\left(\frac{l}{2}!\right)^2
\right\}.~~~~~
\end{eqnarray}
Note that Eq.(\ref{100}) is correct only for even $l>2$, whereas $A_2(l)\equiv0$ for odd values of $l$.

The last subcomponent derived in \cite{LEZ1} in the form of single-series representation was
\begin{equation}\label{101}
\psi_{3,0}^{(2c)}(\alpha,\theta)=\sum_{l=0}^\infty P_l(\cos\theta)(\sin \alpha)^l \phi_l(\rho).
\end{equation}
It is the physical solution of the IFFR
\begin{equation}\label{102}
\left(\Lambda^2-21\right)\psi_{3,0}^{(2c)}=h_{3,0}^{(2c)},
\end{equation}
with the rhs
\begin{equation}\label{103}
h_{3,0}^{(2c)}=-\frac{4\xi}{3\sin \alpha}.
\end{equation}
Function $\phi_l(\rho)$ was obtained \cite{LEZ1} in the form
\begin{equation}\label{104}
\phi_l(\rho)=\phi_l^{(p)}(\rho)+c_l v_{3l}(\rho),
\end{equation}
where the closed expressions for the functions $\phi_l^{(p)}(\rho)$ was derived in \cite{LEZ1} (see also, Appendix \ref{SD}), and function $ v_{3l}(\rho)$ is defined by Eq.(\ref{68}).
The problem is that the coefficient $c_l$ was obtained in very complicated integral form.
A simple form of this coefficient can be written as follows:
\begin{equation}\label{105}
c_l=\frac{2(2l+1)-\pi-H_{\frac{l}{2}}+H_{\frac{l-1}{2}}}{6(2l-3)(2l-1)(2l+1)2^l},
\end{equation}
where $H_z$ are the harmonic numbers. Details can be found in Appendix \ref{SD}.

\section{Conclusions}\label{S4}

The individual Fock recurrence relations introduced in \cite{LEZ1} were used to derive the explicit expressions for the components $\psi_{k,0}^{(0)}$ and $\psi_{k,0}^{(k)}$ of the AFC $\psi_{k,0}$. Using the methods described in \cite{LEZ1}, the mentioned above edge components were calculated and presented for $4\leq k\leq8$. However, given that the IFRR for the edge components of the order $k$ contain only the edge components of the lower order, there is no problem to calculate the edge components with arbitrary $k$. Moreover, it was stated that the components $\psi_{k,0}^{(k)}$ are the functions of the hyperspherical angle $\alpha$ only, whereas  the components $\psi_{k,0}^{(0)}$ are the functions of a single variable $\xi$ defined by Eq.(\ref{I3}).

The single-series representation was derived for the subcomponent $\psi_{3,0}^{(2e)}$ missed in \cite{LEZ1}. This subcomponent is the physical solution of the IFRR (\ref{64}) with the rhs of the form (\ref{63}).
The specific coefficient $s_l$ being a part of the mentioned representation was found in a simple explicit form. This coefficient was derived by the proper application of the \emph{Mathematica} operator \textbf{FindSequenceFunction}.
The same method was applied in order to find a simple expressions for coefficients $c_l$ and $A_2(l)$ included into the single-series representations of the subcomponents $\psi_{3,0}^{(2c)}$ and $\psi_{4,1}^{(2d)}$, respectively.
For the latter subcomponent we derive the closed explicit representations through the hypergeometric functions and the elementary functions, as well.

\section{Acknowledgment}
The author acknowledges Prof. Nir Barnea for useful discussions. 
This work was supported by the PAZY Foundation.

\appendix

\section{}\label{SA}

In this Appendix we describe the details of deriving the subcomponent $\psi_{3,0}^{(2e)}$ representing the solution of the IFRR (\ref{64}) with the rhs of the form (\ref{63}). The physical solution of Eq.(\ref{64}) we shall seek in the form of the single-series (\ref{65}).
First, using the Sack representation for $\xi^{-1}$ (see, e.g., \cite{AM1} or \cite{LEZ1}) let us present the rhs of the IFRR (\ref{64}) in the form
\begin{equation}\label{A1}
h_{3,0}^{(2e)}=\sum_{l=0}^\infty P_l(\cos \theta)\left(\sin \alpha\right)^l\textmd{h}_l(\alpha)
%~_2F_1\left(\frac{l}{2}+\frac{1}{4},\frac{l}{2}+\frac{3}{4};l+\frac{3}{2};\sin^2\alpha\right).
\end{equation}
were
\begin{equation}\label{A2}
\textmd{h}_l(\alpha)=-2^{-l}\sin \alpha
~_2F_1\left(\frac{l}{2}+\frac{1}{4},\frac{l}{2}+\frac{3}{4};l+\frac{3}{2};\sin^2\alpha\right)=
-2^{-l}\sin \alpha
\left\{ \begin{array}{c}
\mathlarger{\sec^{2l+1}\left(\frac{\alpha}{2}\right)},~~0\leq\alpha\leq\frac{\pi}{2}\\
\mathlarger{\csc^{2l+1}\left(\frac{\alpha}{2}\right)}.~~\frac{\pi}{2}<\alpha\leq\pi\\
\end{array}\right.
\end{equation}
For $0\leq\rho\leq1$ (see definition (\ref{13})) Eq.(\ref{A2}) reduces to (see Eq.(B3)\cite{LEZ1})
\begin{equation}\label{A3}
h_l(\rho)\equiv\textmd{h}_l(\alpha)=-2^{1-l}\rho(1+\rho^2)^{l-1/2}.
\end{equation}
To derive the function $\lambda_l(\rho)$ one needs to solve Eq.(44)\cite{LEZ1} for $k=3$ which is
\begin{equation}\label{A4}
\left(1+\rho^2\right)^2\lambda_l''(\rho)+2\rho^{-1}\left[1+\rho^2+
l(1-\rho^4)\right]\lambda_l'(\rho)+(3-2l)(2l+7)\lambda_l(\rho)=-h_l(\rho).
\end{equation}
Using the method of variation of parameters, the particular solution of Eq.(\ref{A4}) can be obtained in the form
\begin{equation}\label{A5}
\lambda_l^{(p)}(\rho)=\frac{1}{2l+1}\left[u_{3l}(\rho)\mathcal{V}_{3l}(\rho)-v_{3l}(\rho)\mathcal{U}_{3l}(\rho)\right],
\end{equation}
where the individual solutions $u_{3l}(\rho)$ and $v_{3l}(\rho)$ of the homogeneous equation associated with Eq.(\ref{A4}) are presented by Eqs.(\ref{67}),(\ref{68})
(see also Eqs.(D6),(D7)\cite{LEZ1}), whereas for the integral factors one obtains
\begin{eqnarray}\label{A6}
\mathcal{U}_{3l}(\rho)\equiv\int \frac{u_{3l}(\rho)h_l(\rho)\rho^{2l+2}}{(\rho^2+1)^{2l+3}}d\rho=
~\nonumber~~~~~~~~~~~~~~~~~~~~~~~~~~~~~~~~~~~~~~~~~~~~~~~~~~~~~\\
-\frac{l(l+1)\left[(\rho^2+1)^4\arctan (\rho)+\rho^7-\rho\right]+(l^2-7l-10)\rho^5-(l^2+9l-2)\rho^3}
{2^l(2l-3)(2l-1)(\rho^2+1)^4},~~~~~
\end{eqnarray}
\begin{eqnarray}\label{A7}
\mathcal{V}_{3l}(\rho)\equiv\int \frac{v_{3l}(\rho)h_l(\rho)\rho^{2l+2}}{(\rho^2+1)^{2l+3}}d\rho=
~\nonumber~~~~~~~~~~~~~~~~~~~~~~~~~~~~~~~~~~~~~~~~~~~~~~~~~~~~~~~~~~~~\\
-\left[(-2)^l(l-2)(l-1)(2l+3)(2l+5)\right]^{-1}
%\left\{12-(l-3)(2l-3)\rho^2\left[2l^2+l-7+(l-2)(2l-1)\rho^2\right]B_{-\rho^2}(l+1,-3)-
\left\{12\left[B_{-\rho^2}(l+1,-3)-B_{-\rho^2}(l+1,-4)\right]+
\right.
~~~\nonumber~\\
\left.
(2l-3)\rho^2\left[2l^2+l-7+(l-2)(2l-1)\rho^2\right]\left[(3-l)B_{-\rho^2}(l+1,-3)-4B_{-\rho^2}(l+1,-4)\right]
\right\}
,~~~~
\end{eqnarray}
%where $B_z(a,b)$ is the Euler beta function.
It is seen that expression (\ref{A7}) cannot be applied directly for $l=1,2$. For this values of $l$, one easily obtains the expressions (\ref{71}), (\ref{72}).

It can be verified that $u_{3l}(\rho)$ is singular, whereas $v_{3l}(\rho)$ and the particular solution $\lambda_l^{(p)}(\rho)$ are regular at the point $\rho=0~(\alpha=0)$ for any $l\geq0$. Hence, the physical solution is of the form
\begin{equation}\label{A8}
\lambda_l(\rho)=\lambda_l^{(p)}(\rho)+s_lv_{3l}(\rho),
\end{equation}
where the coefficient $s_l$ can be found by the coupling equation (61)\cite{LEZ1} which for this case becomes
\begin{equation}\label{A9}
\mathcal{Q}_{2l,l}=\frac{2^{2(l+2)}(l+1)!}{\sqrt{\pi}~\Gamma\left(l+\frac{3}{2}\right)}
\int_0^1\left[\lambda_l^{(p)}(\rho)+s_lv_{3l}(\rho)\right]\frac{\rho^{2l+2}}{(\rho^2+1)^{2l+3}}d\rho.
\end{equation}
Here, $\mathcal{Q}_{n,l}$ denotes the unnormalized HH expansion coefficients for subcomponent
\begin{equation}\label{A10}
\psi_{3,0}^{(2e)}(\alpha,\theta)=\sum_{nl}\mathcal{Q}_{n,l} Y_{nl}(\alpha, \theta).
\end{equation}
To derive the closed expression for $\mathcal{Q}_{n,l}$ we first obtain the  unnormalized HH expansion
\begin{equation}\label{A11}
h_{3,0}^{(2e)}(\alpha,\theta)=\sum_{nl}\mathcal{H}_{n,l} Y_{nl}(\alpha, \theta)
\end{equation}
for the rhs of Eq.(\ref{64}), where by definition
\begin{equation}\label{A12}
\mathcal{H}_{n,l}=N_{nl}^2\int h_{3,0}^{(2e)}(\alpha,\theta) Y_{nl}(\alpha, \theta)d\Omega
\end{equation}
with the normalization constant defined by Eq.(\ref{I8}).
Inserting Eqs.(\ref{A1}),(\ref{A2}) and the HH definition (\ref{I7}) into the rhs of Eq.(\ref{A12}), one obtains
\begin{eqnarray}\label{A13}
\mathcal{H}_{n,l}=-\frac{2^{l+2}N_{nl}^2\pi^2}{2l+1}\left[
\int_0^{\pi/2}\sin^{2l+1}\left(\frac{\alpha}{2}\right)\sin^2\alpha~C_{n/2-l}^{(l+1)}(\cos \alpha)d\alpha+
\right.
~\nonumber~~~~~~~~~~~~~~~~~~~~~~~~~~~\\
\left.
\int_{\pi/2}^\pi\cos^{2l+1}\left(\frac{\alpha}{2}\right)\sin^2\alpha~C_{n/2-l}^{(l+1)}(\cos \alpha)d\alpha
\right],~~~~
\end{eqnarray}
where the orthogonality of the Legendre polynomials was used. On the other hand, substitution of expansion (\ref{A10}) into the lhs of Eq.(\ref{64}) yields
\begin{equation}\label{A14}
(\Lambda^2-21)\psi_{3,0}^{(2e)}=\sum_{nl}\mathcal{Q}_{n,l}(n-3)(n+7)Y_{nl}(\alpha,\theta).
\end{equation}
According to Eq.(\ref{64}), the right-hand sides of Eqs.(\ref{A14}) and (\ref{A11}) can be equated, which yields
\begin{equation}\label{A15}
\mathcal{Q}_{n,l}=\frac{\mathcal{H}_{n,l}}{(n-3)(n+7)}.
\end{equation}
Thus, making use of the relation (\ref{A13}) for $n=2l$, one obtains
\begin{equation}\label{A16}
\mathcal{Q}_{2l,l}\equiv\frac{\mathcal{H}_{2l,l}}{(2l-3)(2l+7)}=
-\frac{2^{l+4}(l+1)!}{\sqrt{\pi}(2l-3)(2l+7)\Gamma(l+3/2)}B_\frac{1}{2}\left(l+2,\frac{3}{2}\right).
\end{equation}
Equating the rhs of Eqs.(\ref{A9}) and (\ref{A16}) we find the required coefficient in the form
\begin{equation}\label{A17}
s_l=\left[\mathcal{I}_1(l)-\mathcal{I}_2(l)\right]\mathcal{I}_3^{-1}(l),
\end{equation}
where
\begin{equation}\label{A18}
\mathcal{I}_1(l)=-\frac{2^{-l}}{(2l-3)(2l+7)}B_{\frac{1}{2}}\left(l+2,\frac{3}{2}\right),
\end{equation}
\begin{equation}\label{A19}
\mathcal{I}_2(l)=\int_0^1 \lambda_l^{(p)}(\rho)\frac{\rho^{2l+2}}{(\rho^2+1)^{2l+3}}d\rho,~~~~~~~~~~~~~~~~
\end{equation}
\begin{equation}\label{A20}
\mathcal{I}_3(l)=\int_0^1 v_{3l}(\rho)\frac{\rho^{2l+2}}{(\rho^2+1)^{2l+3}}d\rho=
\frac{2^{-l-3/2}(2l+1)}{(2l+3)(2l+7)}.
\end{equation}
It is seen that Eqs.(\ref{A17})-(\ref{A20}) yield very complicated expression for the coefficient $s_l$ included (see Eq.(\ref{A8})) into %the single-series representation (\ref{65}) for the subcomponent $\psi_{3,0}^{(2e)}$.
representation (\ref{66}) for $\lambda_l(\rho)$.
The most simple method of finding the simplest representation for $s_l$ is the use of the \emph{Mathematica} operator \textbf{FindSequenceFunction}. In particular, using Eqs.(\ref{A17})-(\ref{A20}) we have calculated the coefficient $s_l$ for $3\leq l\leq30$, and found that it has a form
\begin{equation}\label{A21}
s_l=a_l+b_l\pi+c_l\ln 2,
\end{equation}
where $a_l,b_l$ and $c_l$ are rational numbers.
Making use of the calculated sequences for each of the coefficients $a_l,b_l$ and $c_l$, the \emph{Mathematica} operator \textbf{FindSequenceFunction} enables us to find the general forms of these coefficients as functions of $l$. Notice that for a given sequence there is a minimal number of terms to enable  \emph{Mathematica} to find the formula of the general term. In particular, for the coefficients  $b_l,c_l$  these minimal number is 8, whereas for $a_l$  it equals 22. Finally, one obtains
\begin{equation}\label{A22}
s_l=\frac{2^{-l-3}}{(2l-3)(2l-1)(2l+1)}\left[2l(l+1)\left(H_{\frac{l+1}{2}}-H_{\frac{l}{2}}-\pi\right)+2l+3\right\},
\end{equation}
where $H_z$ are the harmonic numbers.

\section{}\label{SB}

To solve the IFRR (\ref{90}) with the rhs of the form (\ref{91}) we used the single-series representation (\ref{92}) for subcomponent $\psi_{4,1}^{(2d)}$.

It was shown in \cite{LEZ1} that function $\tau_l(\rho)$ from Eq.(\ref{92}) represents the physical solution of equation
\begin{equation}\label{B1}
\left(1+\rho^2\right)^2\tau_l''(\rho)+2\rho^{-1}\left[1+\rho^2+l(1-\rho^4)\right]\tau_l'(\rho)+4(2-l)(l+4)\tau_l(\rho)=
-h_l(\rho),
\end{equation}
were (see Eq.(C6) \cite{LEZ1})
\begin{eqnarray}\label{B2}
h_l(\rho)=-\frac{(\pi-2)(\rho+1)\left(\rho^2+1\right)^{l-1}}{3\pi(2l-1)(2l+3)2^{l+1}}\times
\nonumber~~~~~~~~~~~~~~~~~~~~~~~~~~~~~~~~~~~~~~~~~~~~~~~~~\\
\left[\frac{15-4l(l+1)(4l+11)}{(2l-3)(2l+5)\rho}+4l(2l+3)+2\rho+4(l+1)(2l-1)\rho^2+\frac{(2l-1)(4l+5)\rho^3}{2l+5}\right]~~~
\end{eqnarray}
for $0\leq\rho\leq1$.
Method of variation of parameters enables us to obtain the particular solution of Eq.(B1) in the form (\ref{94}), where
\begin{equation}\label{B3}
\mathcal{V}_{4l}(\rho)=\int_0^\rho v_{4l}(t)h_l(t)t^{2l+2}(1+t^2)^{-2l-3}dt,
\end{equation}
\begin{equation}\label{B4}
\mathcal{U}_{4l}(\rho)=\int_1^\rho u_{4l}(t)h_l(t)t^{2l+2}(1+t^2)^{-2l-3}dt,
\end{equation}
and the individual solutions $u_{4l}$ and $v_{4l}$ of the homogeneous equation associated with (\ref{B1}) are defined by (\ref{95}) according to formulas (46)\cite{LEZ1} for $k=4$.

Our aim is to find the closed representations for the integrals (\ref{B3}) and (\ref{B4}) through the special and elementary functions.
To this end, it would be useful, first of all, to express the solutions $u_{4l}(\rho)$ and  $v_{4l}(\rho)$ through the elementary functions.
It is seen from Eq.(\ref{95a}) that for $l\geq3$, we can write down
\begin{equation}\label{B5}
u_{4l}(\rho)=\frac{8(l-3)!(1+\rho^2)^{l+4}}{15\sqrt{\pi}\Gamma(l+1/2)\rho^{2l+1}}
\sum_{m=0}^{l-3}\frac{(-1)^m\Gamma(m+7/2)\Gamma(l-m+1/2)}{m!(l-m-3)!}\rho^{2m}.
\end{equation}
The well-known formula (7.3.1.140)\cite{PRU3} was applied. Notice that the explicit expressions for the particular solutions $\tau_l^{(p)}(\rho)$ with $l=0,1,2$ were presented by Eqs.(C9)-(C11)\cite{LEZ1}.

Solution of the problem for $v_{4l}(\rho)$, defined by Eq.(\ref{95b}), is more complicated. The use of the relation (7.3.1.9) \cite{PRU3} for $m=l-3$, and subsequent application of the transformation (7.3.1.3) \cite{PRU3} yields
\begin{eqnarray}\label{B6}
~_2F_1\left(\frac{7}{2},l+4;l+\frac{3}{2};-\rho^2\right)=
\frac{16\Gamma(l+3/2)}{105\sqrt{\pi}\rho^{2(l-3)}(1+\rho^2)^7}
\sum_{p=0}^{l-3}\frac{(-1)^p}{p!(l-p-3)!}
\nonumber~~~~~~~~~~~~~~~~\\
\times~_2F_1\left(1,l-p+4;\frac{9}{2};\frac{\rho^2}{1+\rho^2}\right)
.~~~~~~~~
\end{eqnarray}
Next step is application of the relation (7.3.1.132) \cite{PRU3} to the Gauss hypergeometric functions in the rhs of Eq.(\ref{B6}). This gives
\begin{eqnarray}\label{B7}
~_2F_1\left(1,l-p+4;\frac{9}{2};\frac{\rho^2}{1+\rho^2}\right)=
\frac{7\Gamma(l-p+1/2)}{2(l-p+3)!}
\bigg \{\frac{15(1+\rho^2)^{l-p+4}}{8\rho^8}\times
~~~~~~~~~~~~~~~\nonumber~~~~~~~~~~~~~~~~\\
\left[\frac{2\rho}{\sqrt{\pi}} \arctan (\rho)-\sum_{m=1}^3\frac{(m-1)!}{\Gamma(m+1/2)}\left(\frac{\rho^2}{1+\rho^2}\right)^m \right]+
\sum_{m=0}^{l-p-1}\frac{(l-p-m+2)!}{\Gamma(l-p-m+1/2)}(1+\rho^2)^{m+1}
\bigg\}.~~~~~~~~
\end{eqnarray}
Inserting Eqs.(\ref{B6})-(\ref{B7}) into Eq.(\ref{95b}), one obtains the following representation for the second solution of the homogeneous equation:
\begin{eqnarray}\label{B8}
v_{4l}(\rho)=
\frac{\Gamma(l+3/2)}{\sqrt{\pi}}\left(\frac{1+\rho^2}{\rho^2}\right)^{l-3}~\sum_{p=0}^{l-3}
\frac{(-1)^p \Gamma(l-p+1/2)}{p!(l-p-3)!(l-p+3)!}
\bigg \{\frac{(1+\rho^2)^{l-p+4}}{\rho^8}\times
~~~~~~~~~~~~~~~\nonumber~~~\\
\left[\frac{2\rho}{\sqrt{\pi}} \arctan (\rho)-\sum_{m=1}^3\frac{(m-1)!}{\Gamma(m+1/2)}\left(\frac{\rho^2}{1+\rho^2}\right)^m \right]+\frac{8}{15}
\sum_{m=0}^{l-p-1}\frac{(l-p-m+2)!}{\Gamma(l-p-m+1/2)}(1+\rho^2)^{m+1}
\bigg\}.~~~~
\end{eqnarray}
To find the analytic representations for the integrals (\ref{B3}) and (\ref{B4}) it is convenient to present the rhs (\ref{B2}) in the compact form
\begin{equation}\label{B9}
h_{l}(\rho)=a_{0l}(\rho+1)(\rho^2+1)^{l-1}\sum_{n=1}^5 a_{nl}\rho^{n-2},
\end{equation}
where the coefficients $a_{nl}$ are defined by Eq.(\ref{98}).
Inserting representations (\ref{B5}) and (\ref{B9}) into the rhs of Eq.(\ref{B4}), and performing the trivial integration, one obtains
\begin{eqnarray}\label{B10}
\mathcal{U}_{4l}(\rho)=a_{0l}\frac{8(l-3)!}{15\sqrt{\pi}\Gamma(l+1/2)}
\sum_{m=0}^{l-3}\frac{\Gamma(m+7/2)\Gamma(l-m+1/2)(-1)^m}{m!(l-m-3)!}\times
~~\nonumber~~~~~~\\
\sum_{n=1}^5 a_{nl}\left(\frac{\rho^{2m+n}-1}{2m+n}+\frac{\rho^{2m+n+1}-1}{2m+n+1}\right).~~~~~~~~~~~~~~~~
\end{eqnarray}
Using Eq.(\ref{95b}) and Eq.(\ref{B9}) one can write down Eq.(\ref{B3}) in the form
\begin{equation}\label{B11}
\mathcal{V}_{4l}(\rho)=a_{0l}\sum_{n=0}^5 b_{n,l}\int_0^\rho t^{2l+n+1}~_2F_1\left(\frac{7}{2},l+4;l+\frac{3}{2};-t^2\right)dt,
\end{equation}
where the coefficients $b_{n,l}$ are defined by Eq.(\ref{99}).
The use of the relation (1.16.1) \cite{PRU3} yields
\begin{equation}\label{B12}
\mathcal{V}_{4l}(\rho)=a_{0l}\sum_{n=0}^5 \left( \frac{b_{n,l}}{2l+n+2}\right) \rho^{2l+n+2}~_3F_2\left(\frac{7}{2},l+4,l+1+\frac{n}{2};l+\frac{3}{2},l+2+\frac{n}{2};-\rho^2\right).~~
\end{equation}
The latter relation gives the representation of the integral (\ref{B3}) through the generalized hypergeometric functions.
To derive representation of the integral (\ref{B3}) through the Gauss hypergeometric functions,
one should first apply the relation (7.4.1.2) \cite{PRU3}. A  subsequent reorganization of summation along with application of the linear transformation (7.3.1.4) \cite{PRU3} gives the required expression (\ref{97}).

Now we shall show that the Gauss hypergeometric functions included into Eq.(\ref{97}) can be expressed through the elementary functions.
The use of the relation (7.3.1.10) \cite{PRU3} and subsequent application of the linear transformation (7.3.1.3) \cite{PRU3} give
\begin{eqnarray}\label{B13}
~_2F_1\left(l-2,m-\frac{3}{2};k+l+\frac{3}{2};-\rho^2\right)=
\frac{\left(-\rho^2\right)^{3-l}\left(k+\frac{9}{2}\right)_{l-3}}{1+\rho^2}\times
~~\nonumber~~~~~~~~~~~~~~~~~~~~~~~~~~~~~~~~~~~~~~~~~~~\\
\sum_{p=0}^{l-3}\frac{(-1)^p}{p!(l-p-3)!}~_2F_1\left(1,k+p+6-m;k+\frac{9}{2};\frac{\rho^2}{1+\rho^2}\right).~~~~~~~~~~~~~~~~~~~~~~
\end{eqnarray}
In its turn, using (7.3.1.10) \cite{PRU3}, and then applying (7.3.1.3) \cite{PRU3}, one obtains
\begin{eqnarray}\label{B14}
~_2F_1\left(l+m-1,m-\frac{3}{2};k+l+m+\frac{5}{2};-\rho^2\right)=
\frac{\left(k+\frac{9}{2}\right)_{l+m-2}}{(1+\rho^2)\left(-\rho^2\right)^{l+m-2}}\times
~~\nonumber~~~~~~~~~~~~~~~~~~~~~~~~~~~~~~\\
\sum_{p=0}^{l+m-2}\frac{(-1)^p}{p!(l+m-p-2)!}~_2F_1\left(1,k+p+6-m;k+\frac{9}{2};\frac{\rho^2}{1+\rho^2}\right).~~~~~~~~~~~~~~~~~~~~~~
\end{eqnarray}
Finally, application of the relation (7.3.1.132) \cite{PRU3} yields:
\begin{eqnarray}\label{B15}
~_2F_1\left(1,k+p+6-m;k+\frac{9}{2};\frac{\rho^2}{1+\rho^2}\right)=
\frac{\Gamma(k+9/2)\Gamma(p-m+5/2)(1+\rho^2)^{k-m+p+6}}{\pi(k-m+p+5)!\rho^{2(k+4)}}\times
~~\nonumber~~~~~~~~~~~~\\
\left[ 2\rho \arctan(\rho)-\sqrt{\pi}\sum_{s=1}^{k+3}\frac{(s-1)!}{\Gamma(s+1/2)}\left(\frac{\rho^2}{1+\rho^2}\right)^s
\right]+
~~~~~~~~~~~~~~~~~~~~~~~~~~~~~~~~~~~~~~~~~~~~~~~\nonumber~~~~~~\\
\frac{2k+7}{2(k-m+p+5)}\sum_{s=0}^{p-m+1}\frac{(-1)^s\left(m-p-\frac{3}{2}\right)_s(1+\rho^2)^{s+1}}
{\left(k-m+p-s+5\right)_s},~~~~~~~~~~~~~~~~~~~~~~
\end{eqnarray}
where $(a)_n$ is the Pochhammer symbol.
Thus, Eqs.(\ref{B13})-(\ref{B15}) together with Eq.(\ref{97}) give the representation of the integral (\ref{B3}) through the rational functions and the arctangent of $\rho$.
The latter result together with Eqs.(\ref{B5}), (\ref{B8}) and (\ref{B10}) give the partial solution $\tau_l^{(p)}(\rho)$ in terms of elementary functions.

\section{}\label{SC}

In the work \cite{LEZ1} the coefficient $A_2(l)$ being a part of the physical solution (\ref{93}), was derived in general but very complicated (integral) form. In particular (see Appendix C \cite{LEZ1}),
\begin{equation}\label{C1}
 A_2(l)=\frac{1}{\mathcal{P}_{2}(l)}\left[\frac{(\pi-2)2^{-3(l+2)}}{3\pi(2l-1)(l-2)(l+4)}\mathcal{P}_{3}(l)-\mathcal{P}_{1}(l)\right],
\end{equation}
where
\begin{equation}\label{C2}
\mathcal{P}_{1}(l)=
\int_0^{1} \tau_l^{(p)}(\rho)\frac{\rho^{2l+2}}{(1+\rho^2)^{2l+3}}d\rho,~~~~~~~~
\end{equation}
\begin{equation}\label{C3}
\mathcal{P}_{2}(l)=
\int_0^{1} v_{4l}(\rho)\frac{\rho^{2l+2}}{(1+\rho^2)^{2l+3}}d\rho=
\frac{\sqrt{\pi}~2^{-2(l+2)}\Gamma\left(l+3/2\right)}{\Gamma(l/2+3)\Gamma(l/2)},~~~~~~~~
\end{equation}
\begin{eqnarray}\label{C4}
\mathcal{P}_{3}(l)=-\frac{2^{l+1}}{(l+3)(2l-3)(2l+3)(2l+5)}\times
\nonumber~~~~~~~~~~~~~~~~~~~~~~~~~~~~~~~~~~~~~~~~~~~~~~~~~~~~~~~~~~~~~~~~~~~~~~~~~~\\
\left\{
\frac{30}{l+1}-\frac{26}{l+2}+13-4l\left[47-2l(2l(l+3)-9)\right]+2^{l+1}\times
\right.
\nonumber~~~~~~~~~~~~~~~~~~~~~~~~~~~~~~~~~~~~~~~~~~~\\
\left.
\left[
(l+1)\left[4l(l(4l+3)-17)+45\right]B_{\frac{1}{2}}\left(l+\frac{3}{2},\frac{1}{2}\right)+
8l\left[l(l(4l(l+4)+3)-56)-62\right]B_{\frac{1}{2}}\left(l+\frac{3}{2},\frac{3}{2}\right)
\right]
\right\},
\nonumber\\
\end{eqnarray}
%It is seen that even $\mathcal{P}_{3}(l)$ can not be called simple. However, the coefficient $\mathcal{P}_{1}(l)$ can be indeed called very complicated.
The \emph{Mathematica} calculation of the coefficients $A_2(l)$ for any integer $l\geq3$ shows that

1) for odd values of $l$ the coefficients $A_2(l)$ equal zero;

2) for even values of $l$ the coefficients $A_2(l)$ are reduced to the form
\begin{equation}\label{C5}
 A_2(l)=\frac{2-\pi}{\pi^2}\mathcal{A}(l),
\end{equation}
with $\mathcal{A}(l)=(a_l+\pi b_l)$ where $a_l$ and $b_l$ are rational numbers.
Using the effective \emph{Mathematica} code, we have calculated the rational numbers $a_l$ and $b_l$  for $l=4$ up to $l=60$ (with step equals 2).
Making use of the \emph{Mathematica} operator \textbf{FindSequenceFunction} it is possible to find the general simple form of the coefficients $a_l$ and $b_l$.
Remind that for a given sequence there is a minimal number of terms to enable \emph{Mathematica} to find the formula of the general term.
In particular, for the coefficients $a_l$ and $b_l$ these minimal numbers are 10 and 26, corresponding to $l=4,6,8,... 22$ and $l=4,6,8,... 54$ , respectively.
Thus, application of the \emph{Mathematica} operator \textbf{FindSequenceFunction} to the sequences mentioned above yields:
\begin{eqnarray}\label{C6}
\mathcal{A}(l)=\frac{\sqrt{\pi}}{360 l (l-2)\Gamma(l+\frac{1}{2})}\times
~~~~~~~~~~~~~~~~~~~~~~~~~~~~~~~~~~~~~~~~~~~~~~~~~~~~~~~~~~~~~~~~\nonumber~\\
\left\{
\frac{\left[l(l+1)(688 l^4+1376 l^3-2480 l^2-3168 l+465)+450\right]\Gamma\left(\frac{l-1}{2}\right)\Gamma\left(\frac{l+1}{2}\right)}
{(2l-3)(2l-1)(2l+1)(2l+3)(2l+5)}-\frac{56}{l-1}\left(\frac{l}{2}!\right)^2
\right\}.~~~
\end{eqnarray}

\section{}\label{SD}

It was derived in \cite{LEZ1} that function $\phi_l(\rho)$ defined by Eq.(\ref{104}) represents the physical solution
of the inhomogeneous differential equation
\begin{equation}\label{D1}
\left(1+\rho^2\right)^2\phi_l''(\rho)+2\rho^{-1}\left[1+\rho^2+l(1-\rho^4)\right]\phi_l'(\rho)+(3-2l)(7+2l)\phi_l(\rho)=-h_l(\rho)
\end{equation}
with
\begin{equation}\label{D2}
h_l(\rho)= \frac{2^{1-l}\left(\rho^2+1\right)^{l+\frac{1}{2}}\left[(1-2l)\rho^2+2l+3\right]}{3(2l-1)(2l+3)\rho}.
\end{equation}
The particular solution $\phi_l^{(p)}$ of the equation (\ref{D1}) was represented in the form (see Eqs.(101)-(104) \cite{LEZ1})
\begin{equation}\label{D3}
\phi_l^{(p)}(\rho)=\frac{2^{-l}\left(\rho^2+1\right)^{l-\frac{3}{2}}}{3(2l-3)(2l-1)(2l+3)(2l+5)}
\left[2f_{1l}(\rho)+\frac{2f_{2l}(\rho)+f_{3l}(\rho)}{2l+1}\right],
\end{equation}
where
\begin{equation}\label{D4}
f_{1l}(\rho)=\left[9-4l(l+2)\right]\rho+\left(13-4l^2\right)\rho^3,~~~~~~~~~~~~~~~~~~~~~~~~~~~~~~~~~~~~
\end{equation}
\begin{equation}\label{D5}
f_{2l}(\rho)=\left[(2l-3)(2l-1)\rho^4+2(2l-3)(2l+5)\rho^2+(2l+3)(2l+5)\right]\arctan(\rho),~~~~~~~~~
\end{equation}
\begin{eqnarray}\label{D6}
f_{3l}(\rho)=-\left[(2l+3)(2l+5)\rho^4+2(2l-3)(2l+5)\rho^2+(2l-3)(2l-1)\right]\times
\nonumber~~~~~~~~~~~~~~~~~~~~\\
%\frac{\rho}{\left(-\rho^2\right)^{l+1}}\left[\ln(1+\rho^2)+\sum_{k=1}^l \frac{\left(-\rho^2\right)^{k}}{k}\right].~~~~~~~
\frac{\rho}{l+1}~_2F_1\left(1,l+1;l+2;-\rho^2\right).~~~~~~~~~~~~~~~~~
\end{eqnarray}
It was shown that the coefficient $c_l$ included into solution (\ref{104}) can be calculated by the formula:
\begin{equation}\label{D7}
c_l=\frac{\mathcal{M}_1(l)-\mathcal{M}_2(l)}{\mathcal{M}_3(l)},
\end{equation}
where
\begin{equation}\label{D8}
\mathcal{M}_1(l)=\frac{2^{-3l-2}l!\sqrt{\pi}}{3(2l-3)(2l-1)(2l+7)\Gamma(l+3/2)}
~_3F_2\left(\frac{2l-1}{4},\frac{2l+1}{4},l+1;l+\frac{3}{2},l+\frac{3}{2};1\right),
\end{equation}
\begin{equation}\label{D9}
\mathcal{M}_3(l)\equiv\int_0^1v_{3l}(\rho)\frac{\rho^{2l+2}}{\left(\rho^2+1\right)^{2l+3}}d \rho=
\frac{2^{-l-\frac{3}{2}}(2l+1)}{(2l+3)(2l+7)}.~~~~~~~~~
\end{equation}
\begin{equation}\label{D10}
\mathcal{M}_2(l)\equiv \int_0^1\phi_{l}^{(p)}(\rho)\frac{\rho^{2l+2}}{\left(\rho^2+1\right)^{2l+3}}d \rho .
\end{equation}
It is seen that according to Eqs.(\ref{D3})-(\ref{D10}) the coefficient $c_l$ is represented by very complicated function of $l$.
However, the \emph{Mathematica} calculations with any given integer $l\geq0$ show that the parameter $c_l$ has a form $c_{0,l}+c_{1,l}\pi+c_{2,l}\ln 2$, where $c_{i,l}~~(i=0,1,2)$ are rational numbers.
Using the Mathematica operator \textbf{FindSequenceFunction}, one obtains the following simple result
\begin{equation}\label{D11}
c_l=\frac{2l+1-(\pi/2)-\Phi(-1,1,l+1)}{3(2l-3)(2l-1)(2l+1)2^l},
\end{equation}
where $\Phi(z,s,a)$ is the Lerch transcendent. Note that for the case under consideration we have
\begin{equation}\label{D12}
\Phi(-1,1,l+1)=\frac{1}{2}\left[\psi\left(\frac{l}{2}+1\right)-\psi\left(\frac{l}{2}+\frac{1}{2}\right)\right]=
\frac{1}{2}\left(H_{\frac{l}{2}}-H_{\frac{l-1}{2}}\right),
\end{equation}
where $\psi(z)$ and $H_z$ are the digamma function and harmonic number, respectively.
The minimal length of a sequence enables the \emph{Mathematica} to find a simple function that yields the sequence $c_{0,l}$ is 22, whereas for $c_{1,l}$  and $c_{2,l}$  it equals 6.

\newpage

\end{document}